\documentclass[10pt,a4paper]{book}
\usepackage{perugia}
\begin{document}
%
%
\newcommand{\ms}{$M_{\odot}$}
\newcommand{\msb}{$M_{\odot}$~}
\newcommand{\al}{$^{26}$Al}
\newcommand{\fe}{$^{60}$Fe}
\newcommand{\be}{$^{10}$Be}
\newcommand{\ca}{$^{41}$Ca}
\newcommand{\mn}{$^{53}$Mn}
\newcommand{\pd}{$^{107}$Pd}
\newcommand{\tc}{$^{99}$Tc}
\newcommand{\ct}{$^{13}$C}

\talktitle{Short-Lived Nuclei in the Galaxy \\
and in the Early Solar System: \\
Crucial Tests Expected from Gamma-ray Line Observations}

\talkauthors{M. Busso \structure{a,b},
             S. Ciprini \structure{a,c}, and
             F. Marcucci \structure{a,c}
            }

\authorstucture[a]{Physics Department and Astronomical Observatory, University of Perugia,
                    via Pascoli, 06123 Perugia, Italy}

\authorstucture[b]{INAF Torino Astronomical Observatory, via Osservatorio 20, 10025, Pino Torinese, Torino, Italy}

\authorstucture[c]{INFN Perugia Section, via Pascoli, 06123 Perugia, Italy}

\shorttitle{Early Solar System Radioactivities: Tests from
$\gamma$-ray Line Observations}

\firstauthor{Busso et al.}

\begin{abstract}
We discuss some open problems in the understanding of the Early
Solar System abundances of short-lived radioactive isotopes, and
the important clarification expected on this matter by precise
measurements of the average galactic abundances of \al $~$and \fe,
through their gamma-ray lines.
\end{abstract}

\section{Introduction}
Important information on nucleosynthesis mechanisms and on the
formation and early evolution of our solar system is today
provided by radioactive isotopes of mean life $\tau$ between 0.1
and 100 Myr, whose presence as \emph{live} species in the solar
system material at the moment of sun's formation has been
ascertained through meteoritic measurements over the last forty
years (see e.g. \cite{zinner02,busso99}, and references therein).
Table 1, adapted from \cite{busso03} presents an inventory on
these nuclei, with the estimates on their early solar system
abundances as known today.
\par The longer-lived isotopes of Table 1, with meanlives in excess of
10Myr, are normally attributed to the continuous, uniform
production through galactic lifetime
\cite{cameronetal93b,wasserburg96}. In this case one can easily
estimate the equilibrium ratio $N_\textrm{R}/N_\textrm{S}$ for the
short-lived radioactive nucleus R with respect to a stable isotope
S produced in the same process. In the simplest possible galactic
model (a closed box, with instantaneous recycling, evolving for a
time duration $T$) this ratio is \cite{schramm70}:
%
\begin{equation} \label{radioactversusstable}
\left[\frac{N_\textrm{R}(T)}{N_{\rm S}(T)}\right]_{UP} \simeq
\frac{P_{\rm R}p(T)\bar\tau_{\rm R}}{P_{\rm S}<p>T}
\end{equation}
%
Here UP means \textit{uniform production} (in the galaxy), $P_{\rm
S}<p>$ is the stellar production rate of the stable nucleus,
expressed as the product of an assumed constant stellar production
factor $P_{\rm S}$ and of the average effective production rate
$<p>$ over $T$ where $p$ is a function of time, $P_{\rm R}$ is the
stellar production factor for the radioactive isotope and $p(T)$
is the effective rate at time $T$ when production ceased. The
early Solar System (hereafter ESS) abundance of radioactive nuclei
not affected by other mechanisms of production simply reflects
their concentration in the interstellar medium at the moment of
the sun's formation, as established by equation (1)
\cite{cameronetal93b,cameron95,wasserburg96,wasserburg98}.
Nevertheless, some shorter-lived nuclei, like \be, \al, \ca, and
\fe~ have ESS concentrations which are too high to be explained in
this way. In the case of \al, for example, the average galactic
concentration, first measured by \cite{mahoney84} with
HEAO\footnote{\texttt{http://heasarc.gsfc.nasa.gov/docs/heao3/heao3.html}},
and then precisely mapped by COMPTEL (see e.g. \cite{diehl95}), is
of the orders of 2-3 solar masses, yielding a ration \al/$^{27}$Al
$\simeq$ 2$-$4 $\times$ 10$^{-6}$. In \cite{lee76,lee77} is
ascertained that the ESS concentration of \al~ was instead of
$\sim$ 5 $\times$ 10$^{-5}$, i.e. larger by at leas an order of
magnitude.
%
\begin{center}
Table 1.~{\bf  Short-Lived Radioactivities in the Early Solar
System} \vspace{0.1cm}
\begin{tabular}{||c|c|c|c|c|c||}
\hline \hline
(R) & (S) & Mean Life & ($N_{\rm R}/N_{\rm S}$)$_{\rm ESS}$ & Place $^{(\textbf{a})}$ & Ref. \\
       &      &    (Myr)  &                   &       &      \\
\hline \hline
  $^{10}$Be & $^{9}$Be  & 2.2 & $5.2 \times 10^{-4}$  & CAI &
  \cite{mckeegan00}
  \\
  $^{26}$Al & $^{27}$Al  & 1.05 & $5.0 \times 10^{-5}$  & CAI & \cite{lee76,lee77} \\
  &   &   &  &  CH & \\
  &   &   &  &  PD & \\
  $^{41}$Ca & $^{40}$Ca  & 0.15  & $1.5 \times 10^{-8}$ & CAI & \cite{srinivasan94,srinivasan96} \\
  $^{53}$Mn & $^{55}$Mn  & 5.3 & $1.4 \times 10^{-5}$ & CAI & \cite{birck85,rotaru92,lugmair98,hutcheon98} \\
  &   &   &   &  PD& \\
  $^{60}$Fe$^{(\textbf{b})}$ & $^{56}$Fe  & 2.2 & $\ge 1.6 \times $10$^{-8}$& PD & \cite{shukolyukov93a,shukolyukov93b,lugmair95}\\
  &   &  &  $\sim$ 10$^{-6}$&  CAI  &  \cite{birk88} \\
  &  &  &  3 10$^{-7}$ $-$ 2 $\times$ 10$^{-6}$ & UOC &\cite{tachibana03a,tachibana03b,mostefaoui03}\\
  $^{107}$Pd & $^{108}$Pd & 9.4 & $\simeq 2.0 \times 10^{-5}$ & PD & \cite{kelly78}\\
  $^{129}$I & $^{127}$I  & 23 & $1.0 \times 10^{-4}$ & CAI & \cite{reynolds60,jeffery61,brazzle99}\\
  &   &  &  &  CH&  \\
  $^{146}$Sm & $^{144}$Sm & 148 & $7.1 \times 10^{-3}$ & PD & \cite{lugmair77,lugmair92}\\
  $^{182}$Hf & $^{180}$Hf  & 13 & $2.0 \times 10^{-4}$ & PD & \cite{harper94,lee95,yin02,klein02}\\
  &   &   &  &  CH & \\
  $^{244}$Pu & $^{232}$Th  & 115 & $3.0 \times 10^{-3}$ & CAI & \cite{rowe65,alexander71} \\
  &   &   &   &  PD &  \\
\hline \hline
\end{tabular}
\end{center}
\vspace{-14pt}
\begin{itemize}
\itemsep 0pt
\footnotesize{
\item $^{(\textbf{a})}$ \emph{Place} indicates the type of sample from
which the data were derived. CAI refers to Calcium-Alluminum
Inclusions; CH refers to chondrites; PD indicates planetary
differentiates. UOC is used for Unequilibrated Ordinary
Chondrites.
\item $^{(\textbf{b})}$ Estimates for different samples are distributed over two
orders of magnitude. However, recent data, derived from UOC cover
only one dex.
}
\end{itemize}
%
%
%
The consequence of this is straightforward: \al, but also other
short-lived nuclei of meanlife below a few Myr, including \ca~ and
\fe, had to be produced locally, immediately before the formation
of our solar system.
\par Two main scenarios have been considered in the literature to
explain this ''local'' production: synthesis though spallation
reactions in the fast winds of the early sun, during its T Tauri
phase, or contamination by a nearby stellar source. Here we
consider these two alternatives (section 2). We also show section
3) how a stellar source is needed for most radioactivities
considered, and how precise $\gamma$-ray line measurements of \al~
and \fe~ could solve the present uncertainties on the type and
mass of the contaminating star.
%
%
\section{Production of radioactive nuclei in T Tauri winds}
%
Models of spallation reactions have suggested that the fast
particles ejected by the Sun in its T Tauri phase may provide a
source for a few radioactive species, especially \be, \al, \ca,
\mn~ \cite{shu97,lee98,gounelle01}. In particular is possible to
produce \be, \al, \ca, and much of the \mn~ through irradiation of
fast particles from the early Sun \cite{gounelle01}. The model
requires that the irradiation takes place in low-energy impulsive
flares in which $^3$He is enhanced by several orders of magnitude
relative to H in the cosmic rays. The low energy of the cosmic
rays and the high abundance of $^3$He allow the authors to pump up
the production of \al; this in turn suppresses production of Be.
However, the energy spectrum and $^3$He abundance required for
this model to work are at the very extreme limits of flares. In
order to make the scheme work also for \ca, an ad-hoc layered
structure of target objects is also required. Moreover, recent
experimental evidence does not support the X-wind scenario. The
presence of \al~ in a highly volatile Na-rich glass in chondrules
is not compatible with this source \cite{russel96} nor is this
compatible with the observation of small \al~ effects in planetary
differentiates \cite{srinivasan94,srinivasan96,marhas00}. After
the discovery by \cite{mckeegan00} of $^{10}$Be, measurements by
\cite{macpherson01} and by \cite{marhas02} provided evidence for
the presence of $^{10}$Be in early meteoritic material that is not
correlated with \al, and in much younger material (e.g. planetary
differentiates), in which no live \al, nor \ca~ was present (see
also \cite{macpherson03}. This implies that $^{10}$Be was formed
after \al~ and \ca~ had decayed. It seems today that the
solar-wind model remains as an open and plausible scenario only
for \be, which in turn is not produced by stellar nucleosynthesis.
%
%
\section{ESS Radioactivities Clarified by $\gamma$-Line Masurements?}
%
At least in the cases of \al~ and \fe, the ongoing galactic
production
 can be tested through observations. Indeed, their decay is accompanied by
observable gamma-ray lines (at 1.809 MeV for \al~ and at 1.173 and
1.332 MeV for \fe, from the decay of the daughter $^{60}$Co).
Measurements were made for these nuclei by $\gamma$-ray line
instruments (e.g.
COMPTEL\footnote{\texttt{http://cossc.gsfc.nasa.gov/comptel/}} on
board of the Compton Gamma Ray Observatory, and
GRIS\footnote{\texttt{http://lheawww.gsfc.nasa.gov/docs/balloon/New\_GRIS\_homepage/gris.html}}
(the balloon-borne Gamma Ray Imaging Spectrometer). These
observations were reviewed in a number of papers (see e.g.
\cite{diehl96,naya98,prantzos96} and references therein). Assuming
a steady-state production, the results suggest an average
abundance of \al~ in the Galaxy of 3.1~$\pm$~0.9 \msb (COMPTEL)
and in the range from 2.6~$\pm$~0.4 to 4.5~$\pm$~0.7 \msb (GRIS).
%
%
%
   \begin{figure*}[t!]
   \centering
   \resizebox{\hsize}{!}{\rotatebox[]{-90}{\includegraphics{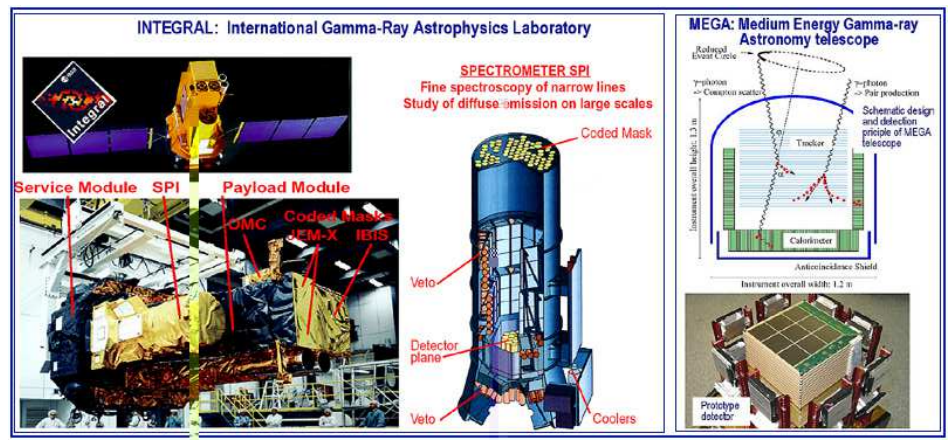}}}
   \vskip -3.6cm
   \caption{\footnotesize{Schematic view of INTEGRAL (left) and MEGA (right) experiments.
\textbf{INTEGRAL}: the ESA International Gamma-Ray Astrophysics
Laboratory is dedicated to the fine spectroscopy ($E/\Delta E =
500$, \textit{SPI} spectrometer) and fine imaging (ang. res.: 12
arcmin FWHM, \textit{IBIS} imager) of celestial gamma-ray sources
in the energy range 15 keV to 10 MeV with concurrent source
monitoring in the X-ray (3-35 keV) and optical (V-band, 550 nm)
energy ranges. (Credits: images based on the Integral Science
Operations Centre. Mission funded by ESA member states,
(especially the PI countries: Denmark, France, Germany, Italy,
Switzerland, Spain), Czech Republic and Poland, with Russia and
USA participation). \textbf{MEGA}: the Medium Energy Gamma-ray
Astronomy project is a MeV space telescope, for gamma-ray imaging
in the energy range from 400keV to 50MeV, and an essential
component of the next generation fleet of gamma-ray space
observatories, filling the gap between hard X-rays/low gamma rays
(INTEGRAL) and high energy gamma-ray missions (AGILE, GLAST). The
project follows in the footsteps of COMPTEL on CGRO, and  is led
by the group of Max-Planck-Institut f\"{u}r Extraterrestrische
Physik (Images reproduced with the permission of G. Kanbach,
MPE).}}
              \label{fig:integral&mega}
    \end{figure*}
%
This average derives from a clumpy distribution, indicative of a
dominant role by massive stars. Recent stellar models suggest a
total production of 2.2~$\pm$~0.4 \msb as coming from massive
stars \cite{knodlseder}. The theoretical and observational
estimates are only marginally compatible, and a further 1 \msb of
\al~ from undetected, low-efficiency or dispersed sources, cannot
be excluded: this might actually improve the agreement
\cite{lentz99}. Though most galactic \al~ certainly comes from
massive stars, they might not be sufficient to account for the
whole galactic inventory, and some contribution might come from
AGB stars \cite{busso99,mowlavi00,nollett03} and novae
\cite{starrfield00,iliadis02}. Similarly, the ESS inventory of
\al~ might have been produced by a low mass AGB star
\cite{busso03}. The nuclide \fe~ has not been detected in the
interstellar medium so far, neither by COMPTEL nor by GRIS.
Massive star nucleosynthesis predicts this nuclide to come from
almost the same stellar zones as \al, and its yield to be in many
cases smaller by only a factor 2--3 (see, e.g..
\cite{meyer00,rauscher02}). A rough rule of thumb would say that,
if only massive stars produce the galactic abundance of both
nuclei, the similar yields and the longer meanlife of \fe~ should
guarantee that the $\gamma$-ray flux for this isotope be rather
close to the present detection limits, whenever \al~ is detected.
However, the available data provide only an upper limit to the
flux ratio, yielding \fe/\al~ $\lesssim$ 0.14 \cite{naya98}. This
is well below the value expected on average from recent SNII
models ($\sim$ 0.6: see \cite{rauscher02}). A low mass AGB star
would produce \al/\fe~ ratios below 0.05, while uncertainties in
ESS measurements still allow the large range \al/\fe~ = (0.02 $-$
0.4). It is therefore evident that new measurements from
$\gamma$-ray lines, positively and quantitatively establishing the
presence of \fe~ in the interstellar medium and its abundance
ratio to \al~ would: i) clarify if recent SN models require
revisions to account for a low \fe~ abundance, as hinted by
present upper bounds; ii) clarify if the ESS \al/\fe~ ratio is
compatible with a typical SN production, or is more similar to low
mass star abundance yields; and iii) establish if point sources of
low intrinsic flux at 1.809 MeV exist, which would confirm the
contribution from (so far undetected) low mass stars. This last
possibility would also give support to the models attributing ESS
short-lived nuclei to a pollution from a nearby low mass star. The
new measurements might be already feasible by the fine gamma-ray
spectroscopy of
INTEGRAL\footnote{\texttt{http://astro.estec.esa.nl/Integral/}}
SPI spectrometer, and will be certainly well within the
possibilities of imaging by
MEGA\footnote{\texttt{http://www.mpe.mpg.de/gamma/instruments/mega/}}
(Fig. \ref{fig:integral&mega}).
%
%
%
%
%

%
%
%
\end{document}